\documentclass[aps,prl,twocolumn,superscriptaddress,showpacs]{revtex4}%
\usepackage{graphicx}
\usepackage{amsmath}
\usepackage{amssymb}
\usepackage{bm}
\usepackage{epsfig}
\begin{document}
\title{\mbox{}\\[10pt]
Fragmentation Contributions to $\bm{J/\psi}$ Production at the Tevatron 
and the LHC}

\author{Geoffrey~T.~Bodwin}
\affiliation{High Energy Physics Division, Argonne National Laboratory,\\
9700 South Cass Avenue, Argonne, Illinois 60439, USA}
\author{Hee~Sok~Chung}
\affiliation{High Energy Physics Division, Argonne National Laboratory,\\
9700 South Cass Avenue, Argonne, Illinois 60439, USA}
\affiliation{Department of Physics, Korea University, Seoul 136-701, Korea}
\author{U-Rae~Kim}
\affiliation{High Energy Physics Division, Argonne National Laboratory,\\
9700 South Cass Avenue, Argonne, Illinois 60439, USA}
\affiliation{Department of Physics, Korea University, Seoul 136-701, Korea}
\author{Jungil~Lee}
\affiliation{High Energy Physics Division, Argonne National Laboratory,\\
9700 South Cass Avenue, Argonne, Illinois 60439, USA}
\affiliation{Department of Physics, Korea University, Seoul 136-701, Korea}

\date{\today}
\begin{abstract}
We compute leading-power fragmentation corrections to $J/\psi$
production at the Tevatron and the LHC. We find that, when these
corrections are combined with perturbative corrections through
next-to-leading order in the strong coupling constant
$\alpha_s$, we obtain a good fit to high-$p_T$
cross section data from the CDF and CMS Collaborations. The fitted
long-distance matrix elements lead to predictions of near-zero $J/\psi$
polarization in the helicity frame at large $p_T$.
\end{abstract}
\maketitle
Much of the current phenomenology of heavy-quarkonium production in
high-energy collisions is based on the effective field theory
nonrelativistic QCD (NRQCD) \cite{Caswell:1985ui}. Specifically,
calculations are based on the NRQCD factorization conjecture
\cite{Bodwin:1994jh}, which states that the inclusive cross
section to produce a quarkonium state $H$ at large momentum transfer
in a collision of particles $A$ and $B$ can be expressed as
\begin{equation}
d \sigma_{A+B \to H + X}= \sum_n d \sigma_{A+B \to Q \bar Q(n) + 
X} \langle  {\cal O}^H(n) \rangle.
\label{NRQCD-fact}
\end{equation}
Here, the $d \sigma_{A+B \to Q \bar Q(n) + X}$ are perturbatively
calculable short-distance coefficients (SDCs),  which are,
essentially, the partonic cross sections to produce a
heavy-quark-antiquark pair $Q\bar{Q}(n)$ in a particular color and
angular-momentum state $n$, convolved with parton distributions. The
$\langle {\cal O}^H(n)\rangle$ are nonperturbative, long-distance
matrix elements (LDMEs) of NRQCD operators and are, essentially, the
probabilities for the pair $Q\bar{Q}(n)$ to evolve into a quarkonium
state $H$ plus anything. The LDMEs are conjectured to be universal,
{\it i.e.}, process independent. This conjecture implies that
information that is gained about the LDMEs by studying one quarkonium
production process can be used to make predictions about another.

The LDMEs have a well-defined scaling with the relative velocity $v$ of
the $Q$ and the $\bar Q$ in the quarkonium center-of-momentum frame.
Consequently, the sum over $n$ in Eq.~(\ref{NRQCD-fact}) is actually
an expansion in powers of $v$, where $v^2\approx 0.25$ for the $J/\psi$
charm-anticharm ($c\bar c$) state. In present-day quarkonium-production
phenomenology, the sum over $n$ is usually truncated at relative
order $v^4$. Four $Q\bar Q$ states appear in this truncation: $Q\bar
Q({}^3S_1^{[1]})$, $Q\bar Q({}^1S_0^{[8]})$, $Q\bar Q({}^3S_1^{[8]})$,
and $Q\bar Q({}^3P_J^{[8]})$, where we use standard spectroscopic
notation for the angular momentum, and the superscripts $1$ and $8$
denote color-singlet and color-octet states, respectively.

Three groups have now completed the formidable task of calculating the
SDCs that appear in the four-LDME truncation through next-to-leading
order (NLO) in the QCD coupling $\alpha_s$
\cite{Gong:2008sn,Gong:2008hk, Gong:2008ft, 
Butenschoen:2010rq,Ma:2010yw,Ma:2010jj,Gong:2012ug}. Generally, the NLO
calculations, combined with the four-LDME phenomenology, lead to
reasonable agreement with a wide range of inclusive
$J/\psi$ production measurements that have been made at the
Tevatron, the LHC, and the $B$ factories \cite{Butenschoen:2011yh,
Butenschoen:2012qr}. Problematic exceptions to this agreement
arise from NLO predictions, which are based on fits to $J/\psi$ cross
sections, that the $J/\psi$ polarization in the helicity frame is
substantially transverse at large $J/\psi$ transverse momentum $p_T$
\cite{Butenschoen:2012px,Butenschoen:2012qr,Gong:2012ug}. Measurements
of the $J/\psi$ polarization at the Tevatron \cite{Affolder:2000nn,
Abulencia:2007us} and the LHC \cite{Chatrchyan:2013cla,Aaij:2013nlm} are
in contradiction with these predictions \cite{footinbib}.

In this Letter we make use of the leading-power (LP) factorization
formalism to compute fragmentation contributions to $J/\psi$ production
beyond NLO that appear in the leading power of $m_{J/\psi}^2/p_T^2$ for
$p_T\gg m_{J/\psi}\approx 2m_c$, where $m_{J/\psi}$ is the $J/\psi$ mass
and $m_c$ is the charm-quark mass. Specifically, we include
contributions that arise from parton production cross sections (PPCSs),
computed through order $\alpha_s^3$ (NLO), convolved with fragmentation
functions (FFs) for a single parton to fragment into a $J/\psi$,
computed through order $\alpha_s^2$ and computed to all orders in
$\alpha_s$ for leading logarithms of $p_T^2/(2m_c)^2$. This procedure
reproduces only part of the full next-to-next-to-leading-order NRQCD
contribution in the large-$p_T$ limit. However, within the LP factorization
framework, it is consistent to calculate the PPCSs and the FFs
separately to a given accuracy, as we do here. We neglect contributions
from the ${}^3S_1^{[1]}$ channel, since they have been found to be small
through NLO in $\alpha_s$ \cite{Gong:2008hk,Gong:2008sn, Gong:2008ft,
Gong:2012ug, Butenschoen:2010rq,Ma:2010yw,Ma:2010jj}. We find that the
new LP fragmentation contributions that we compute have important
effects on the shapes of the SDCs as functions of $p_T$ and on the
relative contributions of the various angular-momentum channels in fits
to the experimental $p_T$ spectra.  We are able to obtain good fits to
the data of the CDF \cite{Acosta:2004yw} and CMS \cite{Chatrchyan:2011kc}
Collaborations for prompt $J/\psi$ production for $p_T\ge 10$~GeV. The
resulting LDMEs lead to a prediction that the $J/\psi$ polarization in
the helicity frame in direct production at both the Tevatron and the LHC
is near zero at high $p_T$, in good agreement with the CMS data
\cite{Chatrchyan:2013cla} and in greatly improved agreement with the CDF
data \cite{Affolder:2000nn, Abulencia:2007us}.

LP factorization states that the contribution to the
inclusive cross section to produce a hadron $H$ at LP in $1/p_T^2$
($d\sigma/dp_T^2\sim 1/p_T^4$) can be written as 
\begin{equation}
d \sigma_{A+B \to H + X}^{\textrm{LP}} 
= \sum_i d\hat\sigma_{A+B\to i+X}\otimes D_{i\to H},
\label{1-ple-frag}
\end{equation}
where the $d\hat\sigma_{A+B\to i+X}$ are inclusive PPCSs to produce a
parton $i$, the $D_{i\to H}$ are FFs
\cite{Collins:1981uw} for the parton $i$ to fragment into the hadron
$H$, and $\otimes$ denotes a convolution with respect to the
longitudinal momentum fraction $z$ of the hadron relative to the
fragmenting parton. The PPCSs can be computed in QCD perturbation
theory; the FFs are nonperturbative and must be determined
phenomenologically. The LP factorization formula in
Eq.~(\ref{1-ple-frag}) was proven for 
the inclusive production of a light hadron in $e^+ e^-$ annililation
in Ref.~\cite{Collins:1981uw}. The proof of
Eq.~(\ref{1-ple-frag}) for production of a heavy quarkonium was
sketched in Ref.~\cite{Nayak:2005rt}. For $J/\psi$ production, the
corrections to Eq.~(\ref{1-ple-frag}) are of relative order
$m_c^2/p_T^2$. Expressions for next-to-leading-power (NLP)
factorization for quarkonium production have been derived in
Refs.~\cite{Kang:2011zza,Kang:2011mg}.

One can apply LP factorization to the SDCs in
Eq.~(\ref{NRQCD-fact}). The result is
\begin{equation}
d \sigma_{A+B \to Q \bar Q(n) + X}^{\textrm{LP}}
=\sum_i d\hat\sigma_{A+B\to i+X}
\otimes D_{i\to Q \bar Q(n)},
\label{sigma-LP-fact}
\end{equation}
where both the PPCSs  $d\hat{\sigma}_{A+B\to i+X}$ and the FFs
$D_{i\to Q \bar Q(n)}$ can be calculated in QCD perturbation
theory. In this Letter, we use the LP factorization approximation 
(\ref{sigma-LP-fact}) for the SDCs to
compute contributions that augment the NLO calculations of the SDCs. As
we have mentioned, we compute the PPCSs $d\hat\sigma_{A+B\to i+X}$
through order $\alpha_s^3$ (NLO), and we compute the FFs $D_{i\to Q \bar
Q(n)}$ through order $\alpha_s^2$ and, for leading logarithms of
$p_T^2/(2m_c)^2$, to all orders in $\alpha_s$.

Formulas for the PPCSs through order $\alpha_s^3$ were given in
Refs.~\cite{Aversa:1988vb,Jager:2002xm}. We evaluate them by making
use of a computer code that was provided by the authors of
Ref.~\cite{Aversa:1988vb}. The gluon FFs $D_{g\to Q\bar Q(n)}$ in
Eq.~(\ref{sigma-LP-fact}) are given for the ${}^1S_0^{[8]}$ channel at
order $\alpha_s^2$ (LO) in Refs.~\cite{Braaten:1996rp,Bodwin:2012xc},
for the ${}^3S_1^{[8]}$ channel at order $\alpha_s$ (LO) in
Ref.~\cite{Braaten:1994kd} and at order $\alpha_s^2$ (NLO) in
Refs.~\cite{Braaten:2000pc,Ma:2013yla}, and for the ${}^3P_J^{[8]}$
channels at order $\alpha_s^2$ (LO) in
Refs.~\cite{Braaten:1994kd,Bodwin:2012xc}. The light-quark FF $D_{q\to
Q\bar{Q}(n)}$ in the ${}^3S_1^{[8]}$ channel is given at order
$\alpha_s^2$ (LO) in Ref.~\cite{Ma:1995vi}. Light-quark fragmentation in
the other color-octet channels vanishes through order $\alpha_s^2$. 

We calculate, to all orders in $\alpha_s$, contributions to the FFs from
leading logarithms of $p_T^2/(2m_c)^2$ by making use of the LO
Dokshitzer-Gribov-Lipatov-Altarelli-Parisi (DGLAP)
evolution equation \cite{Gribov:1972ri, Lipatov:1974qm,
Dokshitzer:1977sg, Altarelli:1977zs}:
\begin{equation}
\frac{d}{d \log \mu_f^2}               
\begin{pmatrix}     
D_S \\ D_g
\end{pmatrix}
=
\frac{\alpha_s(\mu_f)}{2 \pi}
\begin{pmatrix}
P_{qq} & 2 n_f P_{gq} \\
P_{qg} & P_{gg}
\end{pmatrix}
\otimes
\begin{pmatrix}
D_S \\ D_g
\end{pmatrix},
\label{eq:DGLAP}
\end{equation}
where $D_g=D_{g\to Q\bar Q(n)}$, $D_S=\sum_f [D_{q_f\to Q\bar
Q(n)}+D_{\bar q_f\to Q\bar Q(n)} ]$, $f$ is the light-quark or
light-antiquark flavor, $n_f=3$ is the number of active quark flavors,
the $P_{ij}$'s are the splitting functions for the FFs, and $\mu_f$ is
the factorization scale. We have suppressed the dependence of $D_{i\to
Q\bar Q(n)}$ and $d\hat{\sigma}_{A+B\to i+X}$ on $\mu_f$. We solve
Eq.~(\ref{eq:DGLAP}) by taking a Mellin transform with respect to
$z$, integrating $\mu_f$ from $2m_c$ to 
$m_T\equiv\sqrt{p_T^2+4m_c^2}$ in order to incorporate the
logarithms of $m_T^2/(2m_c)^2\approx p_T^2/(2m_c)^2$, and taking an
inverse Mellin transform in order to obtain a $z$-space expression.

\begin{figure}
\epsfig{file=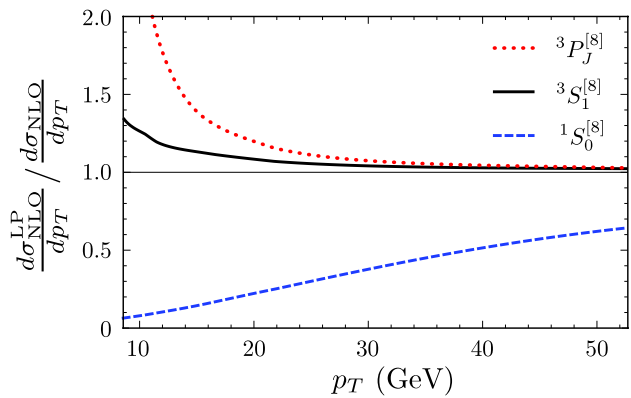,width=8.25cm}
\caption{\label{fig:NLO-frag}%
The ratio 
$(d\sigma^{\textrm{LP}}_{\rm NLO}/dp_T) / (d\sigma_{\rm NLO}/dp_T)$
for the ${}^1S_0^{[8]}$, ${}^3P_J^{[8]}$, and ${}^3S_1^{[8]}$ channels
in $pp\to J/\psi+X$ at $\sqrt{s}=7\,$TeV.
}
\end{figure}

In order to avoid double counting, we must subtract from $d \sigma_{A+B
\to Q \bar Q(n) + X}^{\textrm{LP}}$ [Eq.~(\ref{sigma-LP-fact})]
contributions through
order $\alpha_s^3$, which also appear in the LO and NLO calculations
of the SDCs. We denote these contributions by $d\sigma^{\textrm{LP}}_{\rm
NLO}/dp_T$, and we denote the sum of the LO and NLO contributions to the
SDCs by $d\sigma_{\rm NLO}/dp_T$. The contributions of
$d\sigma^{\textrm{LP}}_{\rm NLO}/dp_T$ to $J/\psi$ production at the LHC
at the center-of-momentum energy $\sqrt{s}=7$~TeV are compared with
$d\sigma_{\rm NLO}/dp_T$  in Fig.~\ref{fig:NLO-frag}. We have taken
$d\sigma_{\rm NLO}/dp_T$ from the calculation of
Refs.~\cite{Ma:2010yw,Ma:2010jj}. In order to maintain compatibility
with that calculation, we have 
taken $m_c=1.5\pm 0.1$~GeV; used the CTEQ6M parton distributions
\cite{Pumplin:2002vw} and the two-loop expression for $\alpha_s$, with
$n_f=5$ quark flavors and $\Lambda_{\rm QCD}^{(5)}=226$~MeV; 
set the renormalization, factorization, and the NRQCD scales to
$\mu_r=m_T$, $\mu_f=m_T$, and $\mu_\Lambda=m_c$, respectively; and 
dropped contributions involving more than one heavy-quark-antiquark
pair in the final state.
 
As can be seen from Fig.~\ref{fig:NLO-frag}, in the ${}^3S_1^{[8]}$ and
${}^3P_J^{[8]}$ channels, $d\sigma^{\textrm{LP}}_{\rm NLO}/dp_T$ accounts
well for the full fixed-order cross section $d\sigma_{\rm NLO}/dp_T$ for
$p_T$ greater than 10--20~GeV. However, in the ${}^1S_0^{[8]}$ channel,
$d\sigma^{\textrm{LP}}_{\rm NLO}/dp_T$ does not approach $d\sigma_{\rm
NLO}/dp_T$ until much larger values of $p_T$. This is a consequence
of the fact that the LP FFs in the ${}^3S_1^{[8]}$ and ${}^3P_J^{[8]}$
channels contain $\delta$ functions and plus distributions
(remnants of canceled infrared divergences)
that are strongly peaked near
$z=1$, while the LP FF in the ${}^1S_0^{[8]}$ channel contains no such
peaking \cite{Bodwin:2012xc}. The NLO correction in the
${}^3S_1^{[8]}$ channel is small relative to the LO contribution because
of a cancellation between the NLO parton-scattering contribution and the
NLO FF contribution, which contribute about $-50\%$ and $+100\%$,
respectively, relative to the LO contribution at $p_T=52.7$~GeV.

\begin{figure}
\epsfig{file=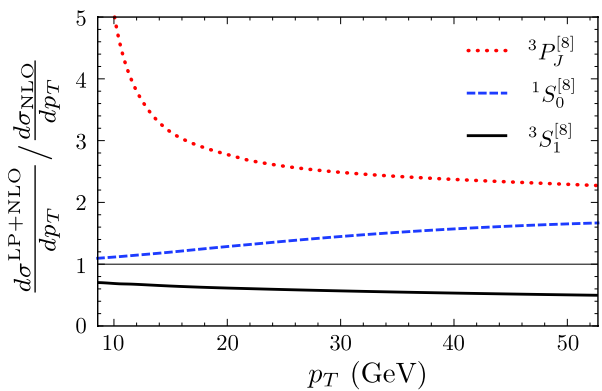,width=7.89cm}
\caption{\label{fig:resum-NLO}%
The ratio 
$(d\sigma^{\textrm{LP$+$NLO}}/dp_T) / (d\sigma_{\rm NLO}/dp_T)$
for the ${}^1S_0^{[8]}$, ${}^3P_J^{[8]}$, and ${}^3S_1^{[8]}$ channels 
in $pp\to J/\psi+X$ at $\sqrt{s}=7\,$TeV.
}
\end{figure}

Our result for the LO plus NLO PPCSs, augmented by the LP 
fragmentation contributions that we have computed, is given by 
\begin{equation}
\frac{d\sigma^{\textrm{LP$+$NLO}}}{dp_T}
=\frac{d\sigma^{\rm LP}}{dp_T}
-\frac{d\sigma^{\textrm{LP}}_{\rm NLO}}{dp_T}
+\frac{d\sigma_{\rm NLO}}{dp_T},
\end{equation}
where $d\sigma^{\rm LP}/dp_T$ is the LP fragmentation
contribution computed to the accuracy described above. In
Fig.~\ref{fig:resum-NLO}, we compare $d\sigma^{\textrm{LP$+$NLO}}/dp_T$ with
$d\sigma_{\rm NLO}/dp_T$ in each channel. The LP corrections in the
${}^3S_1^{[8]}$ and ${}^1S_0^{[8]}$ channels grow in magnitude with
increasing $p_T$, reaching $-50\%$ and $70\%$, respectively, at
$p_T=50$~GeV. The LP corrections are quite dramatic in the
${}^3P_J^{[8]}$ channel, partly because the LO and NLO contributions
tend to cancel at low $p_T$. $d\sigma^{\textrm{LP$+$NLO}}/dp_T$ is 
$80\%$--$160\%$
larger than $d\sigma^{\rm{LP}}_{\rm NLO}/dp_T$ in this channel. These
large corrections suggest that the perturbation expansion may be
converging slowly.

The contributions that we have calculated are dominated by $gg$ initial
states, which account for about $70$\% of the cross section at
$p_T=52.7$~GeV. Contributions from light-quark fragmentation and from
$q$-$g$ mixing in the DGLAP equation amount to only about 1\% and less
than 1\% of the cross section, respectively, at $p_T=52.7$~GeV.

At $p_T=52.7$~GeV, the all-orders resummation of leading logarithms
contributes about $-43$\% in the ${}^3S_1^{[8]}$ channel relative to the
LO fragmentation contribution. Essentially all of that is already
accounted for in the NLO contribution. In the  ${}^1S_0^{[8]}$ and
${}^3P_J^{[8]}$ channels, the all-orders resummations contribute only
2\% and 5\%, respectively, relative to the NLO fragmentation
contribution, owing to an accidental cancellation between contributions
from the running of $\alpha_s$ and contributions from the DGLAP
splitting. Hence, in each channel, $d\sigma^{\rm LP}/dp_T
-d\sigma^{\rm{LP}}_{\rm NLO}/dp_T$  is given to good approximation by
the contribution from the NLO PPCSs convolved with the
order-$\alpha_s^2$ contribution to the FF.

If we vary $\mu_r$ and $\mu_f$ separately between $2m_T$ and
$m_T/2$, then half of the difference between the maximum and minimum values
of the SDCs is less than about 25\% of the central value over the $p_T$
range that we consider. These relatively small scale variations suggest
that perturbation series may be under reasonable control. Overall
factors of $m_c$ in the SDCs can be absorbed into redefinitions of the
LDMEs, and, hence, the uncertainty in $m_c$ from these factors does not
affect fits to the cross sections or the polarization predictions that
we make. The residual $p_T$-dependent effects from the uncertainty in
$m_c$ are less than about 5\%. Therefore, in fitting the data, we
assume that the theoretical uncertainty is 25\%. This value is also
typical of the uncertainty that one would expect from corrections of
higher order in $v$.

\begin{figure}
\epsfig{file=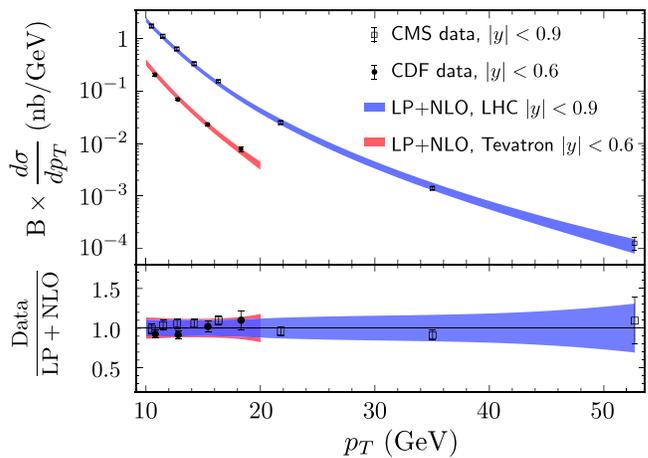,width=8.5cm}
\caption{\label{fig:fits}%
LP$+$NLO predictions for the $J/\psi$ differential cross section at the LHC and
Tevatron compared with the CMS~\cite{Chatrchyan:2011kc} and CDF
data~\cite{Acosta:2004yw}. ${\rm B} = 5.93 \times 10^{-2}$ is the
branching ratio for $J/\psi \to \mu^+ \mu^-$~\cite{Beringer:1900zz}.
}
\end{figure}

In Fig.~\ref{fig:fits}, we show a combined fit of our cross section
predictions to CDF \cite{Acosta:2004yw} and CMS \cite{Chatrchyan:2011kc}
data for prompt $J/\psi$ production. In obtaining these fits, we have
included only data with $p_T\geq 10$~GeV in order to suppress NLP 
corrections. The resulting fit is quite
good, with $\chi^2/{\rm d.o.f.}=0.085$, suggesting that higher-order 
corrections do not affect the $p_T$ dependences of the SDCs at the level of 
our 25\% estimate of the theoretical uncertainty. The fit leads to the
following values for the LDMEs: $\langle {\cal
O}^{J/\psi}({}^1S_0^{[8]})\rangle=0.099\pm0.022$~GeV${}^3$, $\langle
{\cal O}^{J/\psi}({}^3S_1^{[8]})\rangle=0.011\pm0.010$~GeV${}^3$, and
$\langle {\cal
O}^{J/\psi}({}^3P_0^{[8]})\rangle=0.011\pm0.010$~GeV${}^5$. The
corrections that we have computed result in very similar shapes at large
$p_T$ for the SDCs for the ${}^3S_1^{[8]}$ and ${}^3P_J^{[8]}$ channels.
As $p_T$ increases, these SDCs fall much more slowly than do the
experimental data. On the other hand, the contribution of the
${}^1S_0^{[8]}$ channel, including the corrections that we have
computed, matches the shape of the experimental data quite well at large
$p_T$. Consequently, in fits to the experimental data with $p_T\geq
10$~GeV, the sum of the contributions of the ${}^3S_1^{[8]}$ and
${}^3P_J^{[8]}$ channels tends to be small, with the predominant
contribution coming from the ${}^1S_0^{[8]}$ channel. While the LDMEs
$\langle {\cal O}^{J/\psi}({}^3S_1^{[8]})\rangle$ and $\langle {\cal
O}^{J/\psi}({}^3P_0^{[8]})\rangle$ are separately poorly determined, a
full covariance analysis shows that the sum of their contributions is
constrained to be much less than the ${}^1S_0^{[8]}$ contribution. A
fit to the CDF and CMS production cross sections that makes use of the
NLO SDCs without the LP fragmentation corrections also describes the
data well, but does not constrain any of the LDMEs. That fit yields
$\langle {\cal
O}^{J/\psi}({}^1S_0^{[8]})\rangle=-0.030\pm0.381$~GeV${}^3$, $\langle
{\cal O}^{J/\psi}({}^3S_1^{[8]})\rangle=0.023\pm0.057$~GeV${}^3$, and
$\langle {\cal
O}^{J/\psi}({}^3P_0^{[8]})\rangle=0.043\pm0.106$~GeV${}^5$, with
$\chi^2/{\rm d.o.f.}=0.239$.

\begin{figure}
\epsfig{file=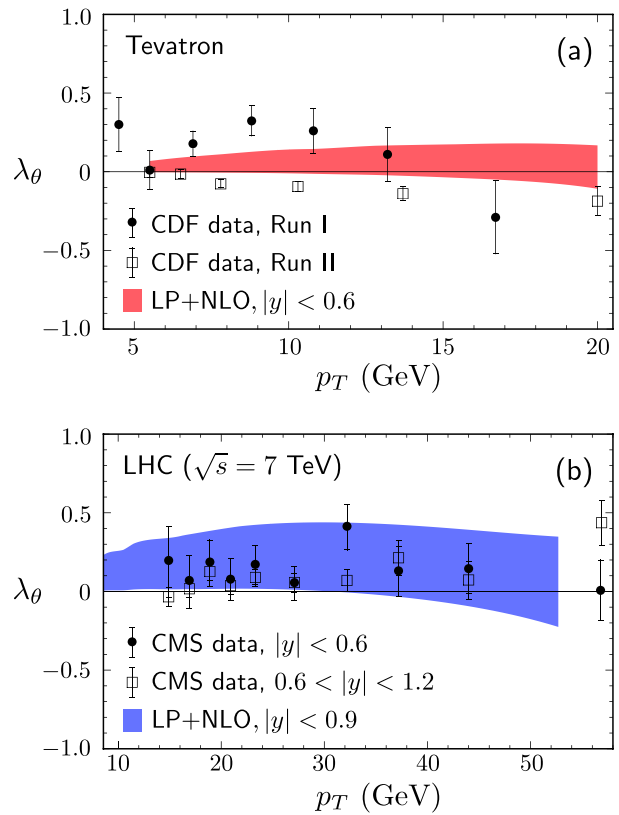,width=8cm}
\caption{\label{fig:pol}%
LP$+$NLO predictions for the $J/\psi$ polarization parameter
$\lambda_\theta \equiv (\sigma_T - 2 \sigma_L)/(\sigma_T + 2 \sigma_L)$
compared with (a) the CDF Run~I \cite{Affolder:2000nn} and CDF Run~II
\cite{Abulencia:2007us} data and (b) the CMS \cite{Chatrchyan:2013cla}
data. Here, $\sigma_T$ ($\sigma_L$) is the cross section for
transversely (longitudinally) polarized $J/\psi$'s.
}
\end{figure}

At high $p_T$, the ${}^3S_1^{[8]}$ and ${}^3P_{J}^{[8]}$ contributions
are both nearly $100\%$ transversely polarized. Hence, the small size of
the sum of the ${}^3S_1^{[8]}$ and ${}^3P_{J}^{[8]}$ contributions
implies that the $J/\psi$'s are produced largely unpolarized at high
$p_T$. (This cancellation of the ${}^3S_1^{[8]}$ and ${}^3P_{J}^{[8]}$
polarization contributions was discussed in Ref.~\cite{Chao:2012iv}, in
which CDF polarization data were used to constrain the LDME fit.) Assuming that
the ${}^3S_1^{[8]}$ and ${}^3P_{J}^{[8]}$ contributions are $100\%$
transversely polarized, we obtain the polarization predictions that are
shown in Fig.~\ref{fig:pol}. These predictions agree with the CMS
\cite{Chatrchyan:2013cla} and CDF \cite{Affolder:2000nn,
Abulencia:2007us} polarization data much better than do the predictions
from the NLO calculations in
Refs.~\cite{Butenschoen:2012px,Butenschoen:2012qr,Gong:2012ug}. The
predictions in
Refs.~\cite{Butenschoen:2012px,Butenschoen:2012qr,Gong:2012ug} rely on
data at $p_T<10$~GeV to constrain the LDMEs. As we have mentioned, a fit
to $J/\psi$ production cross sections at $p_T\geq 10$~GeV that makes use
of the NLO SDCs does not constrain the LDMEs. Consequently, it does not
yield a definite prediction for the polarization.

The LP-fragmentation corrections to $J/\psi$ production that we have
described in this Letter result in substantial changes to the
predictions of NRQCD factorization for $J/\psi$ production. This initial
investigation suggests that these corrections might resolve the
long-standing conflict between NRQCD factorization predictions for
quarkonium polarizations and the polarization measurements that have
been made in collider experiments. Several {\it caveats} should be
mentioned. First, we are comparing theoretical predictions for direct
$J/\psi$ production with prompt $J/\psi$ production data that include
feed down from the $\chi_{cJ}$ and $\psi(2S)$ states. Collider
experiments have yet to determine whether feed-down effects substantially
alter shapes of differential cross sections or measured polarizations.
Second, the large sizes of the corrections that arise from the
parton-scattering cross sections at NLO suggest that the perturbation
expansion may not yet be under good control. Investigations of
higher-order corrections to the PPCSs and FFs should be pursued, as
should NLP fragmentation corrections. Finally, the approach that has
been presented in this Letter should be tested for additional quarkonium
states, such as the $\chi_{cJ}$, $\Upsilon(nS)$, and $\chi_{bJ}$ states,
and for additional production processes.

We thank Jean-Philippe Guillet for providing information about the
computer code that implements the NLO parton-scattering results of
Ref.~\cite{Aversa:1988vb}. We are grateful to Mathias Butensch\"on and
Bernd Kniehl for supplying details of their NLO calculations. We
especially thank Kuang-Ta Chao and Yan-Qing Ma for providing us with
extensive numerical results that are based on their NLO calculations in
Refs.~\cite{Ma:2010yw,Ma:2010jj}. This work was supported in part by
Korea University. The work of G.\,T.\,B. and H.\,S.\,C. is supported by the 
U.S.\ Department of Energy, Division of High Energy Physics, under Contract
No.\ DE-AC02-06CH11357. The submitted manuscript has been created in part by
UChicago Argonne, LLC, Operator of Argonne National Laboratory. Argonne,
a U.S.\ Department of Energy Office of Science laboratory, is operated
under Contract No. DE-AC02-06CH11357. 

\end{document}